\documentclass[
onecolumn,
]{ceurart}

\usepackage{listings}
\usepackage{booktabs}
\usepackage{soul,color}
\usepackage{subcaption}
\usepackage{tabularx}
\usepackage{geometry}
\usepackage{array}
\usepackage{setspace}
\usepackage{hyperref}
\usepackage{fancyvrb}

\begin{document}

\copyrightyear{2024}
\copyrightclause{Copyright for this paper by its authors.
  Use permitted under Creative Commons License Attribution 4.0
  International (CC BY 4.0).}

\conference{ITASEC 2024: Italian Conference on CyberSecurity,
  April 08--11, 2024, Salerno, Italy}

\title{Unveiling Online Conspiracy Theorists: \\a Text-Based Approach and Characterization} 


\author[1]{Alessandra Recordare}[%
email=alessandra.recordare@iit.cnr.it
]
\cormark[1]
\address[1]{Insitute of Informatics and Telematics (IIT), National Research Council (CNR),
  Via G. Moruzzi 1, 56124, Pisa, Italy}

\author[1]{Guglielmo Cola}[%
email=guglielmo.cola@iit.cnr.it
]

\author[1]{Tiziano Fagni}[%
email=tiziano.fagni@iit.cnr.it
]

\author[1]{Maurizio Tesconi}[%
email=maurizio.tesconi@iit.cnr.it
]

\cortext[1]{Corresponding author.}

\begin{abstract}
In today's digital landscape, the proliferation of conspiracy theories within the disinformation ecosystem of online platforms represents a growing concern. This paper delves into the complexities of this phenomenon. We conducted a comprehensive analysis of two distinct X (formerly known as Twitter) datasets: one comprising users with conspiracy theorizing patterns and another made of users lacking such tendencies and thus serving as a control group. The distinguishing factors between these two groups are explored across three dimensions: emotions, idioms, and linguistic features. Our findings reveal marked differences in the lexicon and language adopted by conspiracy theorists with respect to other users. We developed a machine learning classifier capable of identifying users who propagate conspiracy theories based on a rich set of 871 features. The results demonstrate high accuracy, with an average F1 score of 0.88. Moreover, this paper unveils the most discriminating characteristics that define conspiracy theory propagators.
\end{abstract}


\maketitle

\section{Introduction}
\label{sect:introduction}

In the era of social networks, where the proliferation of misinformation and conspiracy theories has become a growing concern, the need to identify users responsible for creating misleading content has become imperative. 
In addressing disinformation, it is crucial to consider the role of social networks, as they have been shown to act as significant amplifiers~\cite{cinelli2020covid}. That is why examining the spread of disinformation within social networks has become an area of growing research interest~\cite{tandoc2019facts, mazza2022, tardelli20}. 
In particular, in the aftermath of the COVID-19 pandemic, there has been an increased focus on the study and understanding of conspiracy theories in general.
This interest stems from the awareness of the significant impact that these theories can have on public health, social cohesion, and the dissemination of accurate information. In response to this challenge, this study aims to provide a contribution by explaining an approach to identifying and profiling individuals who promote conspiracy theories on social media \cite{marcellino2021detecting}.

This study builds upon prior research~\cite{gambini24} where a technique was introduced to collect two datasets: one consisting of apparent conspiracy theorists and the other of generic users, all sourced from X (Twitter). Additionally, in that work a classification study was conducted to differentiate between conspiracy and generic users, using a combination of psycholinguistic features and platform-specific profile characteristics, including Following Count, Follower Count, Bio Sentences, Retweet Ratio, and more. In our research, we seek to characterize conspiracy theorists solely based on their writing style, moving away from dependencies on social network-related features. We explore three distinct categories of features: emotions, idioms, and linguistic attributes.
Moreover, we aim to identify the specific features that prove to be crucial in making this distinction. Given the definition of ``conspiracy user'' as someone who believes in conspiracy theories (conspiracy theorist), our research questions are: 

\textbf{RQ1} -- Is it possible to identify a conspiracy user through text alone? 

\textbf{RQ2} -- What are the features that differentiate a conspiracy user from a generic user?

Our investigation involved training various classifiers using three distinct classes of text-based features, ensuring that our insights could be applied independently of the specific social networking platform.
The results show that the two groups of users exhibit divergent writing styles, underscoring distinct attitudes. Moreover, we identified which types of features are most effective in revealing the tendency to adhere to conspiracy theories.

The paper is organized as follows. In the following Section we briefly present some of the most relevant studies in the field of fake news and conspiracy theories detection. In Section~\ref{sec:dataset}, we describe the X dataset and the preprocessing steps required for our analysis. Next, in Section~\ref{sec:features}, we show and describe the adopted features. Section~\ref{sec:analysis} shows the analysis we performed on the dataset. Finally, in Section~\ref{sec:conclusions}, we summarize our findings and suggest avenues for future research.

\section{Related work}

Social media have greatly facilitated the dissemination of unverified information and misleading content~\cite{cinelli2020covid, mazza2022, cresci14}. There has been extensive research on fake news detection to enable the analysis of fake news spreaders. A relevant example is the study in~\cite{fakenewsdetect}, which revealed that there are characteristics (most dependent on social media) that differ between users who share fake news and those who share real news. Giachanou et al. in \cite{giachanou2020role} proposed a system that exploits a set of psycholinguistic characteristics and personality traits inferred by users to discriminate between potential spreaders of fake news and fact-checkers.

Recent years have seen an increasing focus on the study of conspiracy theories among fake news and disinformation, especially in response to global events such as the COVID-19 pandemic~\cite{cinelli2020covid, zola2022italian}. Alternative explanations for historical or ongoing events, which claim that individuals or groups with malevolent intentions are involved in occult conspiracies, have infiltrated online communication, popular culture, political discourse, and various other areas~\cite{mahl2023conspiracy}.
Researchers are studying how these theories spread across different social platforms, analyzing the mechanisms that lead to their adoption by individuals and trying to identify and characterize conspiracy users in different ways~\cite{mitra2016understanding, memon2020characterizing, schmitz2021python, jimenez2022representational, batzdorfer2022conspiracy, zeng2021conceptualizing}.
While conspiracy theories have recently been associated with vaccines, their scope extends to several other realms.
For example, Marcellino et al. \cite{marcellino2021detecting} collected and analyzed online discussions related to four specific conspiracy theories. 
Klein et al. \cite{c2019pathways} examined users posting a variety of conspiracy theories on Reddit, analyzing differences in the language used by conspiracy theorists compared to other users.
The work of Fong et al. \cite{fong2021language} analyzed conspiracy theories posted by influencer users on Twitter. 
Bessi et al. \cite{bessi2016personality} examined the differences between Facebook users who adhere to conspiracy theories and those who do not, characterizing the personalities of the two groups.

Many of these studies have suggested the possibility of distinguishing the two user groups, yet they lack detailed insights into the extent of this differentiation. Our aim is therefore to distinguish between ``conspiracy users'' and other users on social media and to give them a characterization. Our work differs from those mentioned above in that we determine whether a user is a conspiracy user by using only the text of posted tweets, independently of other dynamics of the social networking platform. In this context, a relevant study is presented in \cite{ConspiDetector}, where a classifier for conspiracy users is described. However, unlike their approach, we do not seek a distinction between users who support conspiracy theories and users who refute them, but instead compare apparent conspiracy theorists with generic users discussing the same topics. Additionally, while their focus was on a narrow range of conspiracies, our study considers a broader set of conspiracy theories.
 
\section{Dataset description}
\label{sec:dataset}

In this section, we present a detailed account of the initial dataset sourced from~\cite{gambini24} as well as the specific preprocessing steps that were executed to adapt the dataset to the objectives of our research.

This initial dataset includes two distinct sets, each consisting of 7,394 X users. The first set, the ``conspiracy group'', comprises users identified as conspiracy theorists. The second set, the ``control group'', includes users not exhibiting apparent conspiracy theory patterns. Conspiracy users were identified by analyzing likes and follows of well-known conspiracy pages or accounts. Instead, the control group consists of users who neither explicitly liked nor followed such content, but still engaged in discussions on the same controversial topics as the conspiracy group and were created around the same time. For each user, the last 3,200 tweets were collected, as of June 13, 2022. The dataset is publicly available\footnote{\url{https://zenodo.org/records/8239530}}.

To ensure the dataset's relevance and reliability for our research objectives, a series of preprocessing steps were undertaken:

\begin{itemize}
\item \textbf{Removal of Retweets}:
To enhance the dataset's suitability for profiling users, we chose to exclude retweets. Retweets, being reposts of other's content, introduce redundancy. By excluding them, we ensured that the dataset primarily consists of original content, aligning with our goal of characterizing users based on their own tweets.

\item \textbf{Language Filter}:
Our analysis focused exclusively on tweets composed in the English language. Implementing this filter was crucial for the subsequent phases of our research and ensured linguistic coherence in our dataset.

\item \textbf{User Tweet Count Threshold}:
To ensure the inclusion of users who have a sufficient presence on X, we implemented a per-user tweet count threshold. Specifically, we excluded users with fewer than 10 tweets within the data collection period. This helped improve the accuracy and reliability of the user profiling we aimed to achieve.

\item \textbf{Selection of Latest 100 Tweets per User}:
Obtaining a large number of tweets from a single user is often challenging in practice. To address this, we focused our analysis on the most recent 100 tweets for each user. This approach reflects more closely real-world scenarios, where several users do not have a high volume of tweet activity.
\end{itemize}

These preprocessing steps transformed the dataset into a more suitable form for our research objectives. As a result, our dataset contained 547,724 tweets from conspiracy users and 592,927 tweets from the control group, posted by a total of 14,568 users. We then balanced the dataset using a Random Undersampling technique, achieving a total of 7,210 conspiracy users and 7,210 control group users.

\section{Method}
The objective of our study is to characterize conspiracy users through a series of steps: identifying suitable features that are dependent solely on the text of the tweet and are not influenced by the platform, conducting a classification task to distinguish between the two groups, and analyzing the most significant features using feature importance metrics. This analysis aims to discern the stylistic differences between the two user groups while ensuring the exclusion of platform-related factors.

In this section we present the features used to characterize users based on their tweets as well as the classification methods employed to discriminate between conspiracy theorists and other users.

\subsection{Features}
\label{sec:features}

We opted to employ three distinct feature groups, all centered around the text content of each individual tweet:

\begin{enumerate}
\item \textbf{Emotions}: We included this feature group to partially implement a sentiment analysis on the dataset. The emotions we have chosen are \textit{Anger}, \textit{Fear}, \textit{Joy}, \textit{Sadness}, \textit{Disgust}, \textit{Surprise}, \textit{Anticipation}, and \textit{Trust}. These emotions align with Robert Plutchik's model of basic emotions, which is widely recognized in the field of psychology \cite{plut}. To assess the emotional content of each tweet, we employed zero-shot learning techniques. Specifically, we used the \texttt{facebook/bart-large-mnli} model available on Hugging Face~\footnote{ \url{https://huggingface.co/facebook/bart-large-mnli}} as a sentiment classifier. This pre-trained model provides a score on a scale from 0 to 1, where a score of 0 indicates no agreement between the emotion and the tweet, while values approaching 1 indicate a strong agreement between them. The facebook/bart-large-mnli model is well known for its accuracy in discerning emotional content in text data \cite{app12178662}. This agreement calculated between the emotion and the tweet will be the feature used for our work.

\item \textbf{Idioms of conspiracy theorists}: 44 sentences were generated by chatGPT-3.5
using the following prompt:
\begin{Verbatim}[fontsize=\small]
What are the typical idioms of a conspiracy theorist? 
Some sayings that come to mind are:
> - think/reason/… with your head
> - they won’t tell you any of this
> - they don’t tell us
> - nobody talks about it
> - wake up!
> - strong powers
> - they make fun of us
> - that’s enough
Do you know any other interesting ones?
\end{Verbatim} 

These idioms are detailed in Table~\ref{tab:idiomi} and they aim to represent the typical language used by conspiracy theorists on social media. The agreement between tweets and idioms was assessed using the zero-shot learning capability of the facebook/bart-large-mnli model, similarly to the method used for emotions. The agreement score, ranging from 0 to 1, was utilized as a feature for subsequent analyses.

\item \textbf{Linguistic features}: We have identified five sets of linguistic features for a total of 72: \textit{lexical} (e.g., num\_words),\textit{syntactical} (e.g. num\_sentences), \textit{semantic} (e.g., num\_named\_entities), \textit{structural} (e.g., avg\_sentence\_length), and \textit{subject-specific} features (e.g., flesch\_reading\_ease). The full list is reported in Table~\ref{tab:linguistic_feat}.

\end{enumerate}

\begin{table}[ht]
  \centering
  \begin{tabular}{|>{\raggedright\arraybackslash}p{0.5\linewidth}|>{\raggedright\arraybackslash}p{0.5\linewidth}|}
    \hline
    \multicolumn{2}{|c|}{\textbf{Idioms}} \\
    \hline
    Behind closed doors  & They want to keep us in the dark \\
    Don't let them catch you & They will not tell you anything about this\\
    Don't let the cat out of the bag & They're cooking up something nefarious\\
    Follow the money & They're out to get us \\
    It's a cover-up & They're planning something behind our backs \\
    It's a deep state conspiracy & They're plotting something sinister \\ 
    It's all part of the plan & They're trying to cover up their tracks  \\
    Nobody talks about it & They're trying to distract us from the real issue \\ 
    Now enough! & They're trying to divide us  \\
    Pulling the strings & They're trying to silence us \\
    Pulling the wool over our eyes & Thinking with your head  \\
    Question everything  & Trust no one \\
    Strong powers  & Wake up!\\
    The conspiracy runs deep &  Watch your back  \\
    The enemy is among us & We have to be prepared \\
    The truth is hidden &  We have to stay one step ahead of them \\
    The truth is out there & We have to stick together\\
    The truth is suppressed & We have to watch our backs \\
    The truth will set us free & We need to be careful who we trust \\ 
    They don't tell us   & We need to dig deeper and uncover the truth  \\
    They don't want us to know the truth  & We need to stay one step ahead of them \\
    They make fun of us  & We need to uncover their secrets \\
    \hline
  \end{tabular}
  \caption{List of idioms used among conspiracy theorists}
  \label{tab:idiomi}
\end{table}

\newcolumntype{L}{>{\raggedright\arraybackslash}X}

\begin{table}[ht]
    \centering
    \begin{tabularx}{\textwidth}{|l|>{\hsize=.7\hsize}L|>{\hsize=.3\hsize}X|}
        \hline
        \textbf{Class} & \textbf{Features}  & \textbf{Description} \\
        \hline
        Lexical & 
        {\scriptsize
        num\_words; num\_unique\_words; num\_chars; num\_unique\_chars; avg\_word\_length; num\_stop\_words; num\_punct; num\_digits; num\_upper\_case\_words; num\_lower\_case\_words; num\_title\_case\_words; num\_proper\_nouns; num\_nouns; num\_verbs; num\_adjectives; num\_adverbs; num\_pronouns; num\_named\_entities; num\_noun\_chunks; num\_exclamation\_marks; num\_question\_marks; num\_spaces} & 
        {\footnotesize
        Word-level characteristics and properties of text. They include various measurements related to the vocabulary and composition of words within a given text.}\\
        \hline
        Syntactical & 
        {\scriptsize
        nominal\_forms; voc\_rich; num\_sentences; avg\_num\_words\_per\_sentence; num\_noun\_phrases; num\_verb\_phrases; num\_adj\_phrases; num\_adv\_phrases; num\_prep\_phrases; num\_coord\_conj; num\_subord\_conj; num\_coord\_clauses; num\_subord\_clauses; punctuation\_freq; num\_capitalized\_sentences; num\_caps\_word\_freq; num\_participial; num\_present\_tense; num\_complementation; num\_relative\_clause} & 
        {\footnotesize
        Grammatical structure and syntax of sentences within a text. They capture the organization and relationships of words and phrases in terms of syntactic rules.}\\
        \hline
        Semantic & 
        {\scriptsize
        num\_personal\_pronouns; num\_impersonal\_pronouns; num\_possessive\_pronouns; num\_reflexive\_pronouns; num\_reciprocal\_pronouns; num\_quantifiers; num\_determiners; num\_prepositions; num\_aux\_verbs; num\_modal\_verbs; num\_negations; num\_synonym; num\_antonymy; 1st\_person\_pronouns; 2nd\_person\_pronouns; num\_passive\_verbs } &  
        {\footnotesize Meaning and interpretation of words and phrases within a text. They capture the underlying semantics and context of language.}\\
        \hline
        Structural & 
        {\scriptsize
        avg\_sentence\_length; avg\_word\_length; avg\_noun\_phrases\_per\_sentence; avg\_verbs\_per\_sentence; proper\_noun\_ratio } & 
        {\footnotesize Overall organization and composition of the text at a higher level, such as sentence and paragraph structure. They provide insights into the textual coherence and complexity.}\\
        \hline
        Subject-specific & 
        {\scriptsize
        flesch\_reading\_ease; smog\_index; flesch\_kincaid\_grade; coleman\_liau\_index; automated\_readability\_index; dale\_chall\_readability\_score; difficult\_words; linsear\_write\_formula; gunning\_fog} & 
        {\footnotesize Specialized indicators relevant to specific domains or topics within the text.}\\
        \hline
    \end{tabularx}
    \caption{List of linguistic features divided by class}
    \label{tab:linguistic_feat}
\end{table}

\begin{table}[ht]
    \centering
    \begin{tabular}{|c|c|c|c|c|c|c|}
        \hline
        \textbf{Emotions} & \textbf{Idioms} & \multicolumn{5}{c|}{\textbf{Linguistic Features}} \\
        \cline{3-7}
        & & Lexical & Syntactical & Semantic & Structural & Subject-specific\\
        \hline
        8 & 44 & 22 & 20 & 16 & 5 & 9 \\
        \hline
    \end{tabular}
    \caption{Number of features used per group}
    \label{tab:num_feat}
\end{table}

Table \ref{tab:num_feat} shows the number of features per each group.

After selecting these features to characterize users, we decided to aggregate tweets by individual users. This transformation resulted in a dataset where each row represents an individual user, rather than an individual tweet. To achieve this, we computed 7 descriptive statistics, namely the mean, median, standard deviation, minimum, maximum, lower quartile, and uppr quartile for the values associated with the tweets of the same user. For instance, when considering the feature 'num\_sentences', the statistics were computed by aggregating values of 'num\_sentences' across all tweets authored by an individual user. 
As a result, each feature was expanded into 7 distinct statistical measures.
This aggregation process provided us with a more holistic perspective on the characteristics of each user, facilitating the summarization of tweet-level attributes into user-level attributes.
After this step, our final dataset comprised 14,420 rows (users) and 868 features.

\subsection{Classification}

The dataset was split into training (85\%) and test (15\%) sets.
Classification was conducted using either the three groups of features individually or a combined set incorporating all of them.
We evaluated a variety of classifiers, including  Logistic Regression, K-Nearest Neighbours (K-NN), Naive Bayes, Support Vector Machine (SVM), Decision Trees, Random Forest, Gradient Boosting, such as XGBoost and LightGBM, Quadratic Discriminant Analysis (QDA), Multilayer Perceptron (MLP), Ridge Classifier, and Linear Discriminant Analysis (LDA). For each classification algorithm, stratified k-fold cross-validation was utilized on the training set to fine-tune the parameters.

\section{Result and Discussion}
\label{sec:analysis}

We first report the results achieved in recognizing conspiracy users from the text contained in their tweets (RQ1). Subsequently, we explore the feature importance within the three groups defined in Section~\ref{sec:features}, in order to unveil the key features that characterize conspiracy theorists (RQ2).

\subsection{Conspiracy users classification}

In the classification task, as mentioned above, we evaluated various classifiers using a single group of features (emotions, idioms, or linguistic) or all of them combined. For emotions, the best performance was achieved with Logistic Regression. For idioms, the best results were obtained through Logistic Regression, Ridge Classifier, and Linear Discriminant Analysis (LDA). On the other hand, for linguistic features and for the combined features, the best performances were achieved using the Light Gradient Boosting Machine (LGBM) algorithm.
Classification results are shown in Table~\ref{class_perf}.

\begin{table}
\centering
\footnotesize
\setlength{\tabcolsep}{4pt}
\begin{tabular}{|l|ccc|ccc|ccc|ccc|}
\hline
 \textbf{Classifier} & \multicolumn{3}{c|}{\textbf{Emotion}} & \multicolumn{3}{c|}{\textbf{Idioms}} & \multicolumn{3}{c|}{\textbf{Linguistic features}} & \multicolumn{3}{c|}{\textbf{All features}}\\
  & \textbf{Prec.} & \textbf{Recall} & \textbf{F1} & \textbf{Prec.} & \textbf{Recall} & \textbf{F1}  & \textbf{Prec.} & \textbf{Recall} & \textbf{F1} &  \textbf{Prec.} & \textbf{Recall} & \textbf{F1} \\
\hline
\textbf{Logistic regression} & \bfseries 0.74 & \bfseries 0.81 & \bfseries 0.77 & \bfseries 0.80 & \bfseries 0.85 & \bfseries 0.82 & 0.79 & 0.83 & 0.81 & 0.83 & 0.85 & 0.84 \\
\textbf{K-NN} & 0.70 & 0.73 & 0.72 & 0.77 & 0.69 & 0.73 & 0.70 & 0.76 & 0.73 & 0.76 & 0.75 & 0.75 \\
\textbf{Naive Bayes} & 0.61 & 0.95 & 0.74 & 0.65 & 0.92 & 0.76 & 0.54 & 0.98 & 0.69 & 0.60 & 0.96 & 0.73\\
\textbf{SVM} & 0.75 & 0.78 & 0.76 & 0.78 & 0.69 & 0.73 & 0.76 & 0.78 & 0.77 & 0.82 & 0.85 & 0.83\\
\textbf{MLP} & 0.73 & 0.74 & 0.73 & 0.81 & 0.80 & 0.81 & 0.80 & 0.78 & 0.79 & 0.85 & 0.84 & 0.85\\
\textbf{Ridge Classifier} & \bfseries 0.73 & \bfseries 0.82 & \bfseries 0.77 & \bfseries 0.80 & \bfseries 0.85 & \bfseries 0.82 & 0.78 & 0.83 & 0.80 & 0.81 & 0.87 & 0.84\\
\textbf{LDA} & \bfseries 0.73 & \bfseries 0.82 & \bfseries 0.77 & \bfseries 0.80 & \bfseries 0.85 & \bfseries 0.82  & 0.78 & 0.83 & 0.80 & 0.81 & 0.87 & 0.84\\
\textbf{DT} & 0.68 & 0.67 & 0.67 & 0.70 & 0.70 & 0.70 & 0.70 & 0.70 & 0.70 & 0.72 & 0.70 & 0.71 \\
\textbf{RF} & 0.73 & 0.79 & 0.76 & 0.76 & 0.84 & 0.79 & 0.77 & 0.77 & 0.77 & 0.78 & 0.84 & 0.81\\
\textbf{XGBoost} & 0.73 & 0.77 & 0.75 & 0.78 & 0.85 & 0.81 & 0.83 & 0.87 & 0.85 & 0.85 & 0.88 & 0.86 \\
\textbf{LightGBM} & 0.74 & 0.79 & 0.76 & 0.78 & 0.85 & 0.81 & \bfseries 0.84 & \bfseries 0.87 &  \bfseries 0.86 & \bfseries 0.86 & \bfseries 0.89 & \bfseries 0.87 \\
\hline
\textbf{Mean} & 0.703 & 0.802 & 0.752 & 0.756 & 0.818 & 0.780  & 0.731 & 0.751 & 0.716 & 0.766 & 0.779 & 0.752 \\
\hline
\end{tabular}
\caption{Precision, Recall and F1 score on test sets}
\label{class_perf}
\end{table}

\subsection{Feature importance}
\begin{figure}
  \centering
  \includegraphics[width=0.55\textwidth]{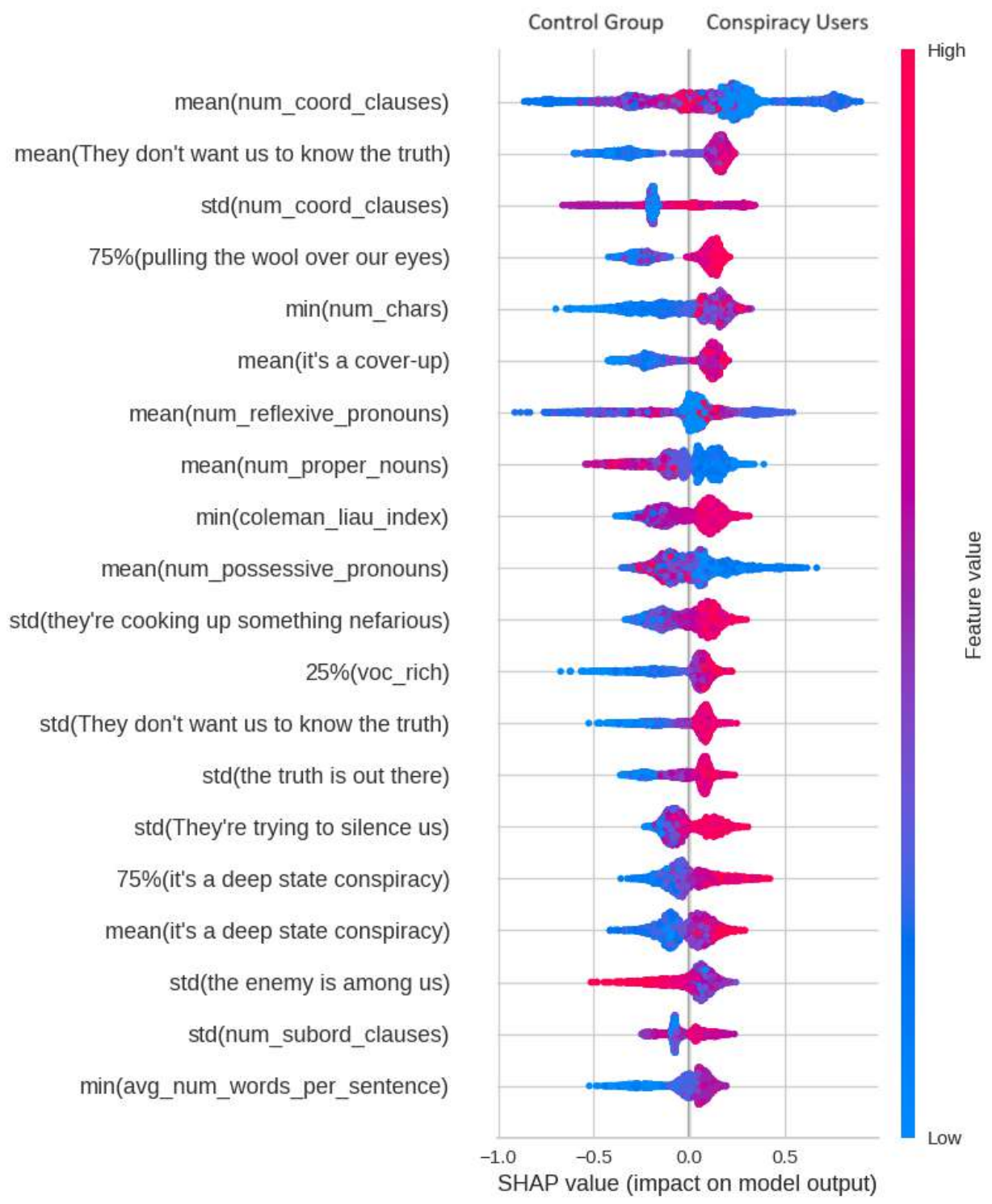}
  \caption{SHAP values for the 20 most important features considering all features}
  \label{fig:shap_all}
\end{figure}

From these results, it is apparent that the feature group which excels at distinguishing between conspiracy users and control group is the set of linguistic features.  Figure~\ref{fig:shap_all} shows the 20 most important features for discriminating between control group users (on the left) and conspiracy users (on the right). Blue points represent low feature values, while red points indicate high values. The SHAP value (the distance from the central vertical axis) indicates the importance of that feature for classification. 
The analysis of the 20 most crucial features for classification shows that the top 10, in terms of importance, originate from the linguistic feature group, with the remaining 10 linked to idioms. Notably, none of the top 20 features are related to emotions, suggesting that emotional features have relatively limited discriminatory power between the two user groups.
The most discriminative feature is \textit{mean(num\_coord\_clauses)}, showing lower values for conspiracy users, followed by \textit{mean(num\_reflexive\_pronouons)} and \textit{mean(num\_possessive\_pronouons)}, both showing higher values for conspiracy users. 

\begin{figure}
  \centering
  \includegraphics[width=0.65\textwidth]{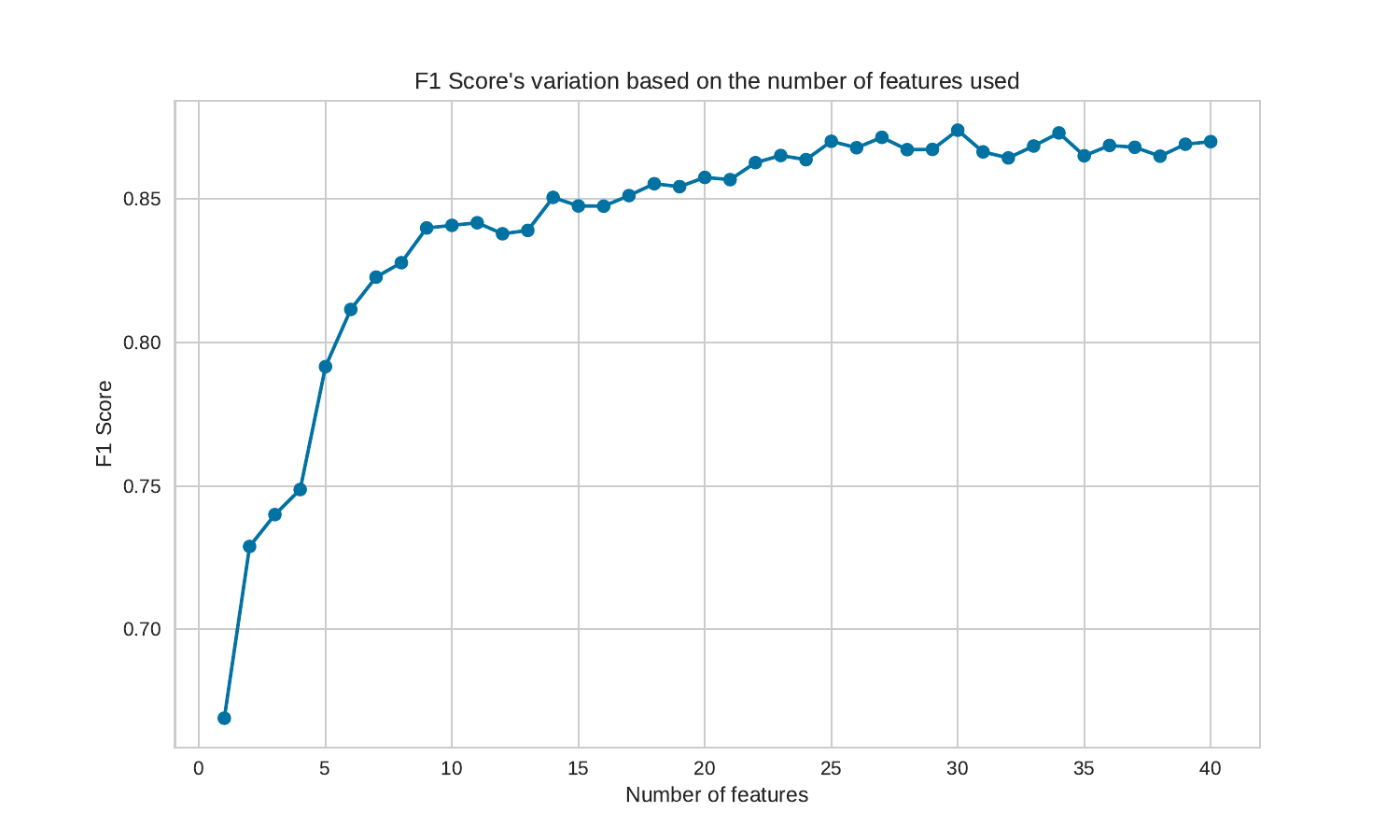}
  \caption{F1 score based on the number of features used for classification (ordered by feature importance) considering all features}
  \label{fig:f1plot}
\end{figure}

Figure~\ref{fig:f1plot} depicts the variation of the F1 score as a function of the number of features employed in the classification, arranged according to their order of importance. We can see that by utilizing the first 30 features, the maximum F1 score is achieved, and notably, even with just the first 14 features, an F1 score of 0.85 is attained.

In the following subsections, we provide a detailed analysis on the relevance of each group of features in recognizing conspiracy users, directly addressing our second research question (RQ2). 

\subsection{Emotions}

\begin{figure}
  \centering
  \includegraphics[width=0.55\textwidth]{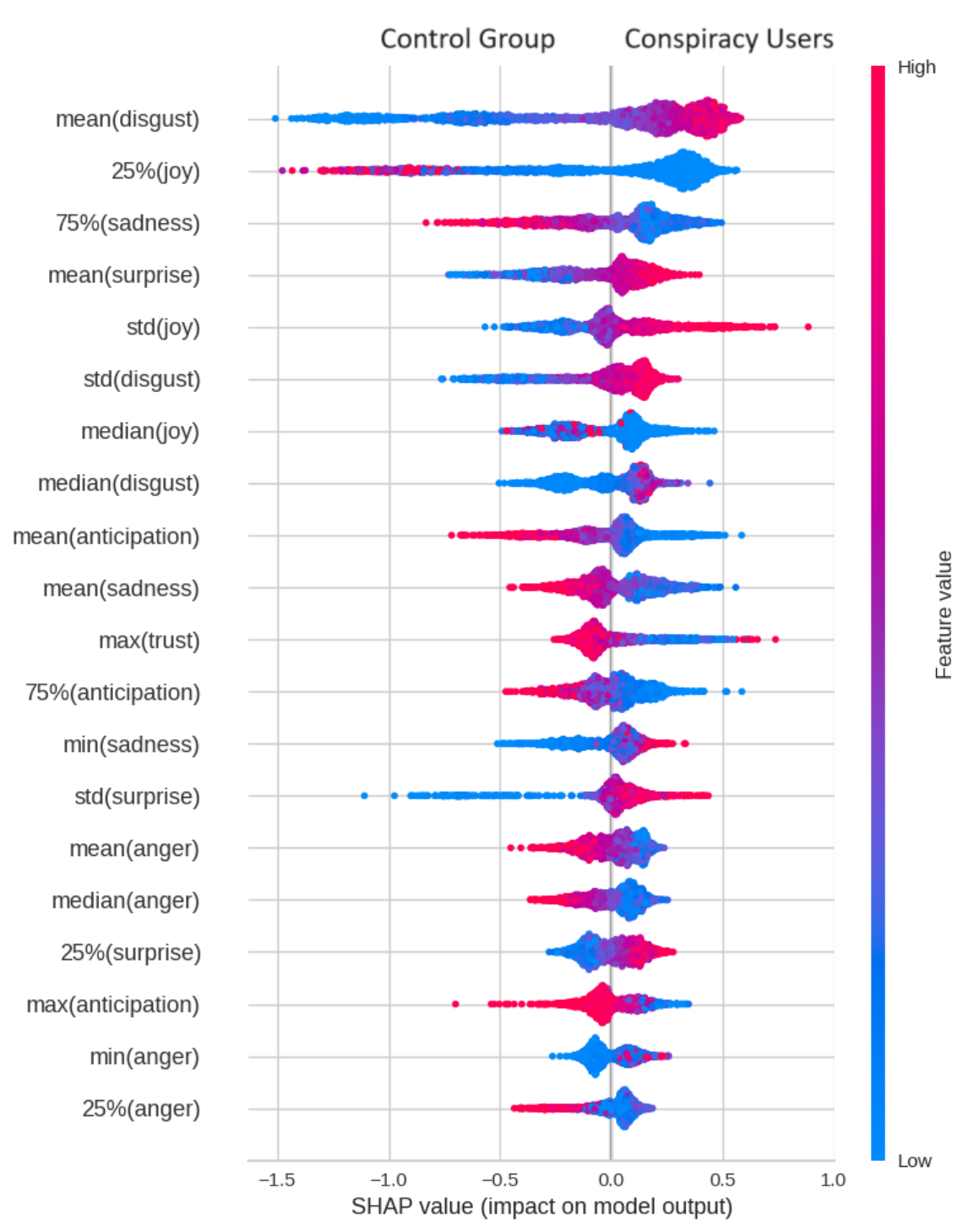}
  \caption{SHAP values for the 20 most important features in the emotions set}
  \label{fig:shap_em}
\end{figure}

We conducted an in-depth analysis of emotion-based feature importance in the LGBM classifier and observed that the most prominent distinguishing emotion between the two user groups is ``disgust'', followed by ``joy'', ``sadness'', and ``anticipation''.
Figure~\ref{fig:shap_em} shows the 20 most important emotional features for discriminating between control group users and conspiracy users. 

For the ``disgust'' emotion, the average, median, standard deviation, and  75th percentile values were significantly higher for conspiracy users. In the control group, the mean and seventy-fifth percentile of the ``joy'' emotion exhibited higher values.
As for the ``sadness'' emotion, conspiracy users showed higher mean and 75th percentile values compared to the control group.
Interestingly, for the ``anger'' emotion, both the mean and median were higher in the control group.

\subsection{Idioms of conspiracy theorists}

\begin{figure}
  \centering
  \includegraphics[width=0.9\textwidth]{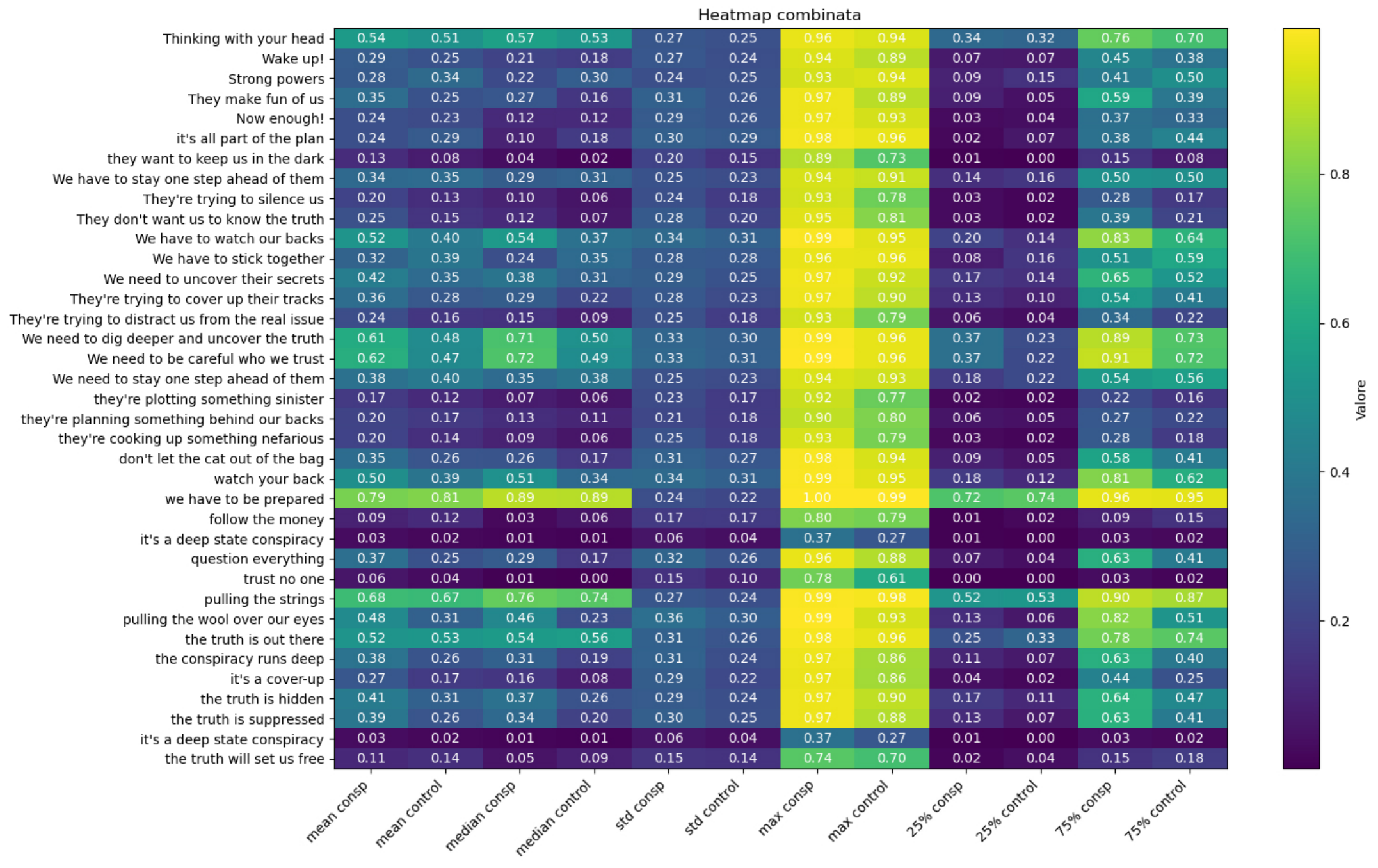}
  \caption{Descriptive statistics relative to the analyzed idioms, for control group users and conspiracy users}
  \label{fig:idioms_}
\end{figure}

Figure~\ref{fig:idioms_} is a heatmap showing the average values of several descriptive statistics
for the majority of idioms in our analysis, divided by user group (control group and conspiracy users). The average of each descriptive statistic was computed among all the users in a group. Cells with higher values tend towards yellow, whereas lower values are represented by violet blue cells.
We excluded the least discriminating idioms and statistics for better readability.

Most of the idioms identified by ChatGPT tend to align more closely, on average, with tweets from conspiracy theorists, except for \textit{We have to stick together}, \textit{Strong powers}, \textit{It's all part of the plan}, \textit{The truth will set us free}, and \textit{Follow the money}, which align more with tweets from the control group.
Some idioms exhibit a strong agreement with both user groups, like \textit{We have to be prepared}, while others show little agreement for either group, such as \textit{Trust no one}, \textit{They're plotting something nefarious}, and \textit{The enemy is among us}.
The average standard deviations are consistently higher for conspiracy theorists, suggesting that this group has more diverse data among themselves.
There are substantial differences in the averages of the 75th percentiles, for example, in phrases like \textit{Pulling the wool over our eyes}, \textit{Question everything}, and \textit{The conspiracy runs deep}, where quite higher values are noted for conspiracy users. This indicates that agreement values for these idioms tend to be higher for this class of users.

\subsection{Linguistic features}
\begin{figure}
  \centering
  \includegraphics[width=0.55\textwidth]{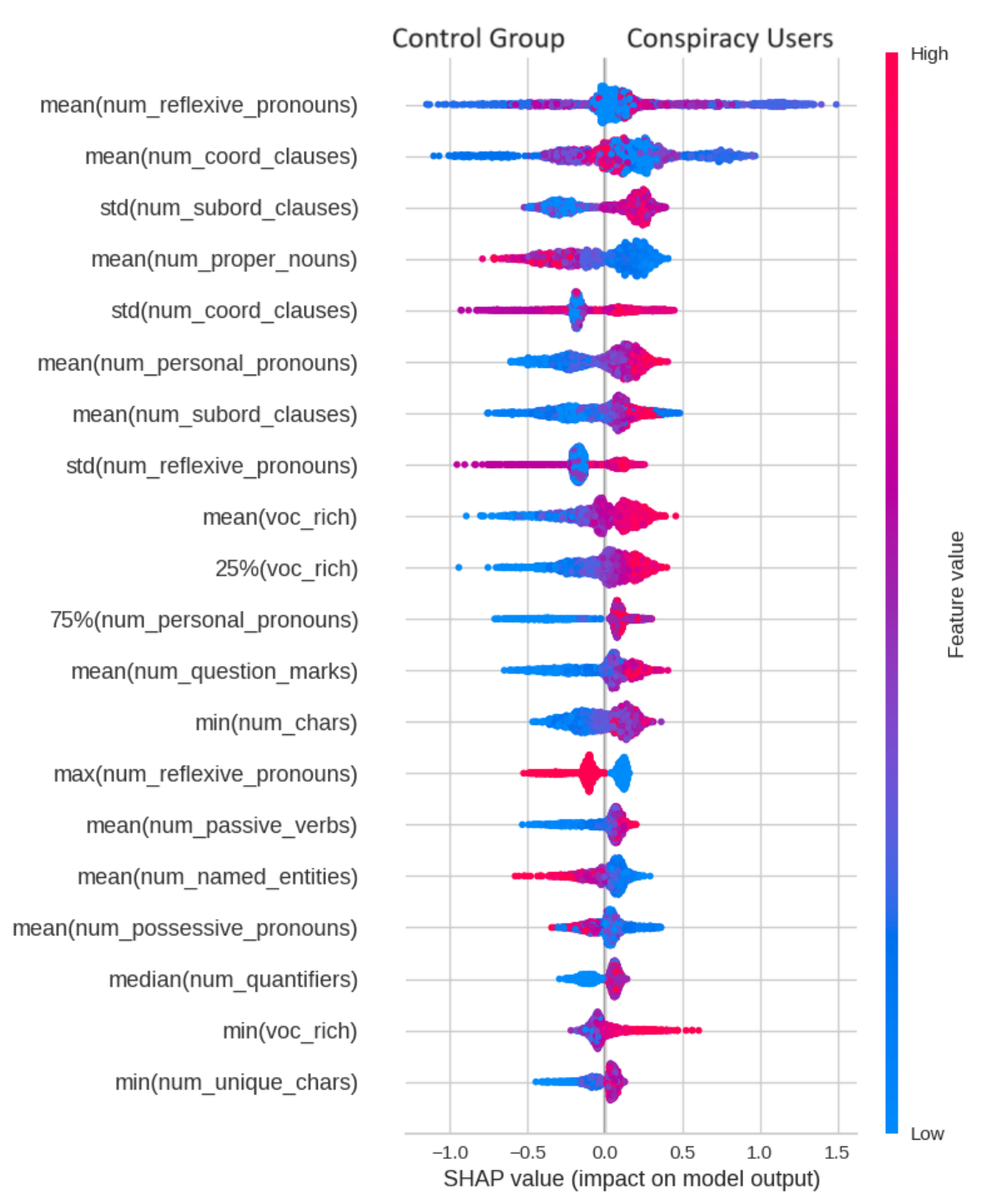}
  \caption{SHAP values for the 20 most important features in the linguistic set}
  \label{fig:shap_feat}
\end{figure}

As previously mentioned, linguistic features have proven to be the most effective in classifying conspiracy and control group users. To further explore this, we divided these features into five groups: lexical, syntactical, semantic, structural, and subject-specific features. Our goal was to ascertain which of these groups contributed most significantly to the differentiation between the two user classes. 
Regarding their utility for classification, we found that semantic features ranked the highest, followed by syntactical, lexical, subject-specific, and, lastly, structural features.
This observation is corroborated by Figure \ref{fig:shap_feat}, which illustrates that the top 20 features contributing to classification predominantly belong to the semantic, syntactical, or lexical categories.

Among the most significant semantic features are the average count of reflexive pronouns, the average count of possessive pronouns (in both cases, the number of pronouns mentioned is higher for conspiracy users), and the average count of named entities (conspiracy users tend to mention fewer entities).
As for syntactical features, the mean and standard deviation of the number of coordinating clauses, along with the standard deviation of the number of subordinate clauses, were identified as the most important. Both of these features exhibited higher values among conspiracy theorists.
Among the prominent lexical features, the mean count of digits (higher for conspiracy users) and the standard deviation of title case word count (also higher for conspiracy users) were found to be the most influential.
Other noteworthy features include the higher count of question marks among conspiracy users, as well as vocabulary richness, indicating a more sophisticated word choice in their tweets. Furthermore, all readability indices suggest a greater level of reading difficulty (and therefore lower readability) in tweets of conspiracy users. This is likely attributed to the usage of acronyms or hashtags typical of the movement they support.

\section{Conclusions and future work}
\label{sec:conclusions}

In this study, we introduced a method for profiling users who endorse conspiracy theories, focusing specifically on characterizing their writing style. This characterization was achieved by analyzing textual content of tweets, intentionally excluding platform-dependent metrics such as likes, retweets, and comments.

We selected a dataset consisting of 14,420 users, evenly split between two categories: 7,210 conspiracy users and 7,210 control group users, who did not exhibit explicit conspiracy theory behavior patterns. For each user, we analyzed between 10 to 100 of their most recent tweets, calculating scores based solely on textual content. These scores were subsequently aggregated for each user, using statistical measures like mean and median to capture the essence of each user's textual patterns.
We also implemented and tested classification algorithms, with the Light Gradient Boosting Machine classifier yielding the most promising results. This classifier enabled us to effectively differentiate between conspiracy and control users, achieving an F1 score of 0.87. 

Responding to RQ1, this research has shown that users can be categorized based solely on the characteristics of their writing style. Furthermore, in response to RQ2, this study identified specific linguistic traits that can be considered characteristic of conspiracy theorists, thus shedding light on the distinct markers of this group within the digital landscape.
We found that the features that best characterize conspiracy users from the control group are linguistic features, in particular the number of coordinate clauses, the number of possessive and reflexive pronouns. Our work shows that conspiracy theorists use fewer coordinate clauses than the control group but more reflexive and possessive pronouns, use more digits, name fewer entities, use a richer vocabulary and have worse readability. Regarding sentiment analysis, the tweets from conspiracy users show a higher agreement with disgust and sadness, while the tweets of the control group are more akin to joy and anger. Considering the set of conspiracy idioms generated via chat-GPT, it turns out that most of them have a higher agreement with conspiracy users.

In future work, we plan to extend our classifier's  application to other platforms like Telegram. This will help assess the model's generalizability and robustness across diverse social media. 
Furthermore, we plan to improve the accuracy of our model by incorporating a wider range of text-only features, enhancing our understanding of user behavior and the overall ability of recognizing conspiracy theorists. 
Additionally, we are interested in exploring different time windows to capture evolving trends and emerging patterns in the propagation of conspiracy theories. 
Lastly, an interesting avenue for future research is examining the implications of our findings on disinformation mitigation strategies. This could lead to more effective methods to counter the spread of disinformation and promote digital literacy.

\begin{acknowledgments}
 We acknowledge the support provided by:
 \begin{itemize}

\item Project SoBigData.it, which receives funding from European
Union – NextGenerationEU – National Recovery and Resilience Plan (Piano Nazionale di Ripresa e Resilienza, PNRR) – Project: "SoBigData.it – Strengthening the Italian RI for Social Mining and Big Data
Analytics" – Prot. IR0000013 – Avviso n. 3264 del 28/12/2021;

\item Project SERICS (PE00000014) under the NRRP MUR program funded by the EU – NGEU;

\item Regione Toscana, which partially funded this work within the framework of the “Por Fse 2014-2020 Investimenti a favore della crescita, dell’occupazione e del futuro dei giovani Fondo SocialeEuropeo”. Decreto di finanziamento DD 21607 del 29/11/2021.

\end{itemize}
  \begin{figure}[h]
    \centering
    \includegraphics[width=0.9\textwidth]{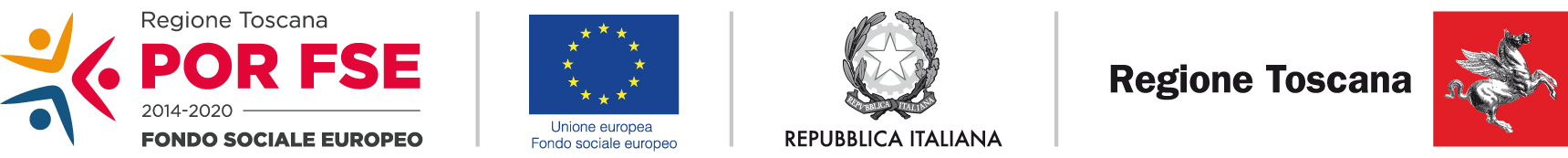} 
\end{figure}

\end{acknowledgments}

\bibliography{references}

\end{document}